%% file: main.tex
\documentclass[conference]{IEEEtran}

\newif\ifCameraReady{}
\CameraReadytrue{}

\input{0_preamble.tex}

\begin{document}

\ifCameraReady{}
\newcommand{\Thanks}{
  This research was supported by DARPA COMPASS HR0011-25-3-0242, DOD MURI grant FA9550-24-1-0327, and DOE ASCR grant DE-SC0023171.
  This work was performed under the auspices of the U.S.\ Department of Energy by Lawrence Livermore National Laboratory under Contract DE-AC52-07NA27344 (LLNL-PROC-2014338),
  Office of Science, Office of Advanced Scientific Computing Research, Scientific Discovery through Advanced Computing (SciDAC) Program through the FASTMath Institute.
  This research used resources from the Swiss National Supercomputing Centre (CSCS) and the National Energy Research Scientific Computing Center (NERSC) under ALCC allocation ERCAP0030671, ScienceAtScale DDR-ERCAP0034808, and NESAP DDR-ERCAP0038013.
}
\fi

\newcommand{\Title}{Accelerating High-Order Finite Element Simulations at Extreme Scale with FP64 Tensor Cores}
\ifCameraReady{}
\title{\Title\\\thanks{\Thanks}}
\else
\title{\Title\vspace{-0.2em}}
\fi

\ifCameraReady{}
\author{
  \IEEEauthorblockN{
    Jiqun Tu\IEEEauthorrefmark{1},
    Ian Karlin\IEEEauthorrefmark{2},
    John Camier\IEEEauthorrefmark{3},
    Veselin Dobrev\IEEEauthorrefmark{3},
    Tzanio Kolev\IEEEauthorrefmark{3},
    Stefan Henneking\IEEEauthorrefmark{4},
    Omar Ghattas\IEEEauthorrefmark{4}
  }
  \IEEEauthorblockA{\IEEEauthorrefmark{1}\textit{NVIDIA Corporation} jtu@nvidia.com}
  \IEEEauthorblockA{\IEEEauthorrefmark{2}\textit{Queen's University} ian.karlin@queensu.ca}
  \IEEEauthorblockA{\IEEEauthorrefmark{3}\textit{Lawrence Livermore National Laboratory} \{camier1, dobrev1, kolev1\}@llnl.gov}
  \IEEEauthorblockA{\IEEEauthorrefmark{4}\textit{The University of Texas at Austin} \{stefan, omar\}@oden.utexas.edu}
} %
\else
\author{}
\fi

\maketitle

\input{1_abstract.tex}

\section{Introduction}\label{sec:introduction}
\input{2_introduction.tex}

\section{Background}\label{sec:background}
\input{3_background.tex}

\section{Performance optimizations}\label{sec:optimizations}
\input{4_optimizations.tex}

\section{Results}\label{sec:results}
\input{5_results.tex}

\section{Related work}\label{sec:related}
\input{6_related.tex}

\section{Conclusions}\label{sec:conclusions}
\input{7_conclusions.tex}

\printbibliography{}

\end{document}

%% file: 0_preamble.tex
\usepackage{amsmath,amssymb,amsfonts}
\usepackage{algorithmic}
\usepackage{graphicx}
\usepackage{textcomp}
\usepackage{xcolor}
\usepackage[table]{xcolor}
\usepackage{physics}
\usepackage{url}
\usepackage{tabularray}
\usepackage{multirow}
\usepackage{xspace}
\usepackage{tikz}
\usetikzlibrary{positioning}
\usetikzlibrary{calc}

\def\BibTeX{{\rm B\kern-.05em{\sc i\kern-.025em b}\kern-.08em
T\kern-.1667em\lower.7ex\hbox{E}\kern-.125emX}}

\usepackage[ruled]{algorithm2e}
\usepackage{booktabs}
\usepackage{mathtools}
\usepackage{comment}

\ifCameraReady
\IEEEoverridecommandlockouts
\fi

\usepackage[backend=biber,
  sorting=none,
  maxbibnames=99,
  maxcitenames=99,
  giveninits=true,
  isbn=false,
  url=false,
  doi=true]
{biblatex}
\usepackage[hidelinks]{hyperref}
\hypersetup{colorlinks=true, citecolor=blue, linkcolor=blue, urlcolor=blue}
\usepackage{cleveref}

\definecolor{RED}{RGB}{190, 30, 49}
\definecolor{BLUE}{RGB}{11, 78, 179}
\definecolor{GREEN}{RGB}{0, 171, 86}
\definecolor{OLIVE}{RGB}{93, 150, 72}
\definecolor{GREY}{RGB}{50, 50, 50}
\definecolor{BROWN}{RGB}{122, 90, 40}

\def\lb{\langle}
\def\rb{\rangle}

\newcommand{\be}{
\begin{equation}}
  \newcommand{\ee}{
\end{equation}}

\newcommand{\bes}{
\begin{equation*}}
  \newcommand{\ees}{
\end{equation*}}

\newcommand{\p}{\partial}

\usepackage{listings}
\makeatletter
\newcommand{\linebreakand}{%
\end{@IEEEauthorhalign}
\hfill\mbox{}\par
\mbox{}\hfill
\begin{@IEEEauthorhalign}
}
\makeatother

\DeclareRobustCommand{\IEEEauthorrefmark}[1]{\raisebox{0pt}[0pt][0pt]{\textsuperscript{\footnotesize #1}}}

%% file: 1_abstract.tex
\begin{abstract}
  Finite element simulations play a critical role in a wide range of applications, from automotive design to tsunami modeling and computational electromagnetics.
  Performing these simulations efficiently at the high resolutions needed for practical applications and scientific insights necessitates the use of high-order methods and large-scale supercomputing.
  While much progress has been made in porting finite element codes to GPU systems in recent years, additional improvements in the efficiency and computational speed of GPU-accelerated high-order finite element simulations are in constant demand.
  In this paper, we demonstrate that the FP64 tensor cores on NVIDIA GPUs can be used to further accelerate such simulations, achieving significant speedups in key kernels of MFEM, a scalable open-source finite element library widely used in HPC applications.
  By integrating FP64 tensor cores with kernel fusion optimizations, we were able to achieve up to 2$\times$ performance gains and up to 83\% energy efficiency gains on NVIDIA's Grace Hopper GH200 and Grace Blackwell GB200 architectures.
  To the best of our knowledge, this is the first time that FP64 tensor cores have been directly programmed to accelerate large-scale finite element scientific computing applications.
  We demonstrate the performance of the optimized kernels at exascale by showing near-perfect weak scaling efficiency and 90\% strong scaling efficiency across nearly 10,000 GPUs on the Alps system.
  The new algorithms and MFEM enhancements directly benefit complex production codes, including the 2025 Gordon Bell Prize-winning application for real-time tsunami forecasting.
\end{abstract}

\begin{IEEEkeywords}
  GPU supercomputing,
  tensor cores,
  DMMA PTX,
  HPC,
  performance optimizations,
  high-order finite elements
\end{IEEEkeywords}

%% file: 2_introduction.tex
\emph{Tensor cores} were introduced in 2017 in the NVIDIA Volta V100 GPU architecture in order to drive higher performance in applications that use dense matrix multiplication.
Initially, the tensor cores supported only mixed-precision (FP16, FP32) \emph{matrix-multiply-accumulate} (MMA) computations where single-precision results are achieved through a combination of reduced-precision (FP16) multiplication and single-precision (FP32) accumulation~\cite[Fig.~9]{nvidia-v100-whitepaper}.
As specialized hardware for MMA instructions, tensor cores delivered dramatically higher throughput for these workloads than standard FP16 and FP32 CUDA cores.
For single-precision performance, tensor cores provided up to 8$\times$ higher FLOP/s performance than CUDA cores on V100~\cite{nvidia-v100}, and the gains have since increased to over 15$\times$ on current Blackwell-class GPUs~\cite{nvidia-gb200} with even larger speedups for lower-precision (FP4, FP8) computations.

While many codes can benefit directly from the performance gains of lower-precision or mixed-precision tensor core computations, there are also many scientific applications that require the full accuracy of double precision computations.
Some examples include singular perturbation problems, solutions with geometric singularities or boundary layers, and multiscale physics problems.
In this paper, we apply tensor cores to one such application: an inverse problem for finite element-discretized wave propagation that aims to infer a high-dimensional parameter field from sparse, noisy data.
This type of inverse problem is ill-posed~\cite{ghattas2021learning}, making the solution sensitive to small errors; stable inversion necessitates double precision calculations.

Starting in 2020 with the Ampere A100 GPU, NVIDIA introduced FP64 tensor cores supporting \emph{native IEEE double precision MMA} (DMMA) instructions.
For double precision calculations, these tensor cores deliver a 2$\times$ increase in peak floating-point rates~\cite{nvidia-a100}.
In recent years, other vendors have introduced tensor cores, or analogous concepts, in their compute architectures.
In this paper, we focus specifically on the programming of tensor cores on two of NVIDIA's state-of-the-art chips: the Grace Hopper GH200 Superchip~\cite{nvidia-gh200} and the Grace Blackwell GB200 Superchip~\cite{nvidia-gb200}.

For applications that use FP64 dense linear algebra---specifically, General Matrix-Matrix Multiplication (GEMM)---for large matrices, including density functional theory, tensor-core usage has led to large performance gains~\cite{yu2022gpu}.
This has been exploited both in scientific applications (e.g., BerkeleyGW~\cite{deslippe2012berkeleygw}) and in benchmarking codes (e.g., HPL~\cite{dongarra1979linpack}) which typically utilize NVIDIA's FP64 tensor cores by calling the GEMM APIs in CUBLAS and other vendor libraries to achieve significant speedups for large GEMMs.
However, many applications, in diverse fields such as molecular dynamics, computational engineering, and weather modeling, have not been able to leverage tensor cores as successfully~\cite{Domke2020MatrixEF} because they either do not perform matrix-matrix multiplies or their operations do not map well to tensor units.
More recently, tensor core computations have been extended to more applications through direct programming, including stencil codes~\cite{gu2025sptcstencil} and finite element methods~\cite{cui2024acceleration}.

We build on this prior work and apply tensor cores to a full finite element production application---a digital twin for tsunami early warning---that is already well-optimized and won the Gordon Bell Prize in 2025~\cite{henneking2025bell}. The application code solves a coupled acoustic--gravity wave propagation problem as a first-order system of partial differential equations (PDEs) discretized with high-order finite elements.
Specifically, we decompose the tensor contractions into small GEMMs of order $\mathcal{O}(10)$ and program the FP64 tensor cores directly in CUDA kernels. We use Parallel Thread Execution (PTX) instructions to directly program tensor cores on state-of-the-art chips like NVIDIA's Grace Hopper GH200 and Grace Blackwell GB200. We also create custom mappings to handle irregular matrix shapes and fuse kernels for additional performance.
This leads to significant performance (up to 2$\times$) and energy efficiency (up to 83\%) gains in a complex, already well-tuned application code.
The resulting application demonstrates near-ideal weak scaling and 90\% strong scaling efficiency on 9,216~GPUs.

Using the tensor-core-optimized kernels, we then demonstrate excellent full-system scalability of the application code on one of the world's largest supercomputers---the \emph{Alps} system at the Swiss National Supercomputing Centre (CSCS).

The main contributions of this paper are:
\begin{itemize}
  \item Detailed design and analysis of the programming and optimization of FP64 tensor cores for irregularly shaped matrix multiplies.
  \item To the best of our knowledge, the first example of using directly programmed FP64 tensor cores in a complex, PDE-based HPC application, in order to improve throughput of high-order finite element kernels (by up to 59\%).
    We show that shared memory data motion, not FLOPs, is the performance bottleneck, and demonstrate strong correlation between tensor core data motion reduction and improved application performance.
  \item Performance comparison of GH200 and GB200 FP64 tensor cores including energy efficiency analysis (27\% improvement), which has not been reported for small matrix tensor core operations before.
  \item Significant finite element kernel performance gains (up to 2$\times$) and energy efficiency gains (up to 83\%) by combining loop fusion techniques with tensor core kernels.
  \item Demonstration of exascale full-system scalability of the application code for each of the implemented kernel versions on a leading-edge supercomputer.
\end{itemize}

The rest of the paper is organized as follows. Section~\ref{sec:background} provides a brief overview of the MFEM library, the finite element algorithms and application used, and the NVIDIA FP64 DMMA tensor cores.
We then discuss how we optimized the key finite element kernels in Section~\ref{sec:optimizations}.
Section~\ref{sec:results} shows the performance and energy efficiency improvements from our optimizations.
In Section~\ref{sec:related} and Section~\ref{sec:conclusions}, respectively, we describe related work and summarize our conclusions.

%% file: 3_background.tex
\subsection{MFEM library}
In order to make our techniques more accessible, we focus our work on finite element kernels in the MFEM library~\cite{mfem-web}, an open-source high-performance finite element library that is widely used in the scientific computing community.
MFEM is designed for solving PDEs via a variety of discretizations on general unstructured grids, with a strong emphasis on high-order methods that enhance accuracy and efficiency by exploiting tensor-product structures~\cite{mfem2021}.
By supporting arbitrary-order spaces and higher-order geometries, MFEM enables simulations across millions of tasks,
incorporating GPU acceleration to tackle the complexities of high-order elements and challenging geometries
while rethinking numerical algorithms to maximize fine-grain parallelism and minimize energy-intensive data movement on exascale architectures.
The library provides a robust foundation for developing custom high-performance solvers,
integrating seamlessly with established libraries like HYPRE~\cite{hypre2002}, SUNDIALS~\cite{hindmarsh2005sundials}, and PETSc~\cite{petsc-user-ref},
as well as specialized multigrid techniques including algebraic multigrid optimized for GPUs.
MFEM's modular architecture empowers diverse applications in computational physics and engineering, including compressible and incompressible flows, electromagnetics,
inertial confinement fusion, additive manufacturing, topology optimization, structural mechanics and medical modeling~\cite{mfem2024}.

\subsection{Finite element assembly algorithms on GPUs}
To leverage GPUs in high-performance computing on exascale platforms, MFEM's finite element algorithms employ tensor factorization strategies, partial assembly (PA, by precomputing quadrature data), and matrix-free (MF, with zero storage but dynamic computations) techniques, in order to expose fine-grain parallelism and reduce data motion.
These on-the-fly operator evaluations are tailored to high-order discretizations, but are generally applicable across all finite element spaces in the \emph{de Rham}~\cite{deRham82} complex.
In addition to PA and MF, MFEM also includes global sparse matrix and element-local dense matrix assembly levels,
to facilitate integration with external solvers and support state-of-the-art simulations for large-scale, complex problems in scientific computing.
\begin{figure*}[htbp]
  \includegraphics[width=\textwidth,trim={0 1.5em 0 1em}]{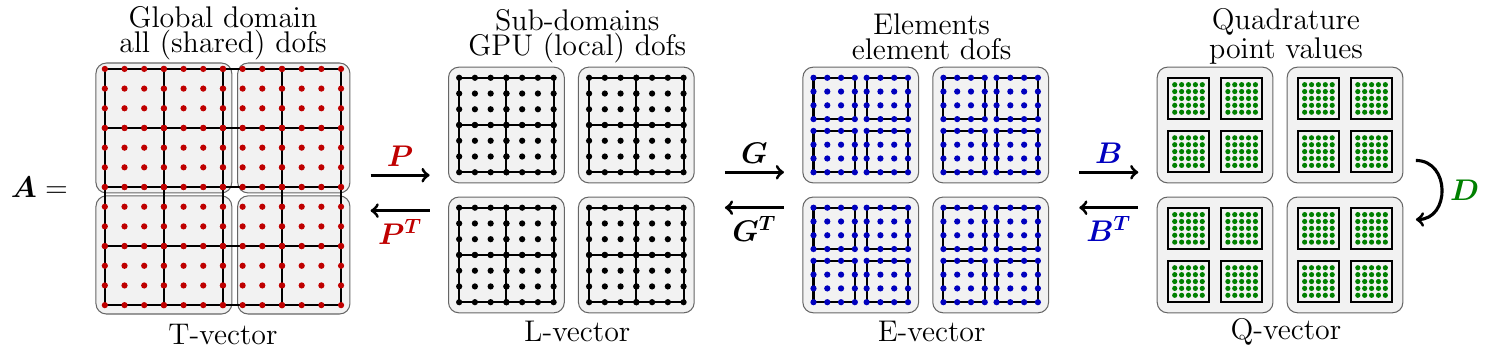}
  \caption{Finite element operators, $A$, where
    $P$ handles parallel scattering and MPI communication across global true degrees of freedom (T-vectors),
    $G$ manages mesh topology to local subdomains (L-vectors) and elements (E-vectors),
    $B$ encodes geometry mappings and tensor-product basis functions,
    and $D$ encapsulates physics at quadrature points (Q-vectors)---with numerical kernels reducing to
  batched small dense tensor contractions implemented as matrix-matrix multiplications (DGEMMs) for performance portability via vendor-optimized BLAS.}
  \label{fig:mfem-operator}
  \vspace{-1.1em}
\end{figure*}

GPU-accelerated PA kernels cover a broad range of operators, enabling end-to-end acceleration of matrix-free solvers through low-order-refined preconditioning~\cite{lor2023}.
Kernel fusion methods enhance strong scalability and allow attaining peak performance on problem sizes 5--10 times smaller than conventional methods.
These matrix-free approaches, distinct from traditional sparse matrix assembly, significantly reduce memory footprint by avoiding global matrix storage,
enhance arithmetic intensity to achieve near-optimal floating-point operations per degree of freedom (DOF),
and minimize memory accesses and global data movement.

One way to describe the PA and MF algorithms is through the decomposition of a finite element operator $A$ into successive applications of simpler operators---$P$, $G$, $B$ and $D$---that represent parallel communication, mesh topology, basis functions, and physics at quadrature points, respectively. This is illustrated in Figure~\ref{fig:mfem-operator}.
\begin{equation}
  \label{eq:mfem-operator}
  A = P^T \, G^T \, B^T \, D \, B \, G \, P
\end{equation}
In the PA approach, we precompute and store the (block-diagonal) matrix $D$ at quadrature points, but all other operators ($P$, $G$ and $B$) are applied on-the-fly.
In contrast, the MF approach applies all parts of the decomposition on-the-fly, including $D$, which reduces storage, but increases FLOPs.
In both cases, we can exploit the tensor-product structure and implement the application of $B$ and $B^T$ in (\ref{eq:mfem-operator}),
which maps DOFs to values at quadrature points and vice versa, through small dense tensor contractions within the 1D version of this mapping, $B^{1d}$.
Since $B_{abc,ijk} = B^{1d}_{ai}B^{1d}_{bj}B^{1d}_{ck}$, a typical pattern for these tensor contractions is:
\begin{equation}
  \label{eq:tensor-notation}
  Y_{abc} = B^{1d}_{ck}B^{1d}_{bj}B^{1d}_{ai}X_{ijk}.
\end{equation}
This approach is known as \emph{sum factorization}.
Combining the PA and MF approaches with sum factorization on tensor-product elements has proven essential for tackling exascale-level problems in MFEM.

A prime example of that is the recent development of a digital twin for tsunami early warning, where MFEM facilitated large-scale simulations of acoustic--gravity wave propagation~\cite{henneking2025bell}.
This work showcased how high-order finite element discretizations can be applied to ill-posed inverse problems in geophysics, setting the stage for the specific PDE model discussed next.

\subsection{Digital twin for tsunami early warning}
Henneking et al.~\cite{henneking2025goal, venkat2025fft} recently developed a novel, scalable framework for real-time Bayesian inference for inverse problems governed by linear time-invariant dynamical systems.
They applied this framework to create a physics-based digital twin for tsunami early warning in the Cascadia subduction zone~\cite{henneking2026cascadia}---a 1000~km long megathrust fault located off the coast of the U.S.\ Pacific Northwest and Canada with the potential for a magnitude 8--9 earthquake and large-scale tsunami.
Given real-time pressure data from seafloor sensors, the digital twin can infer the earthquake-induced seafloor motion and forecast tsunami wave heights in less than a second.
This work won the 2025 ACM Gordon Bell Prize~\cite{henneking2025bell}.

As part of the digital twin creation, a large number---equal to the number of sensor locations informing the tsunami warning (which were assumed to be on the order of hundreds)---of computationally demanding PDE solutions describing the governing tsunami dynamics had to be computed in an offline precomputation step.
This precomputation constituted by far the most expensive computational phase of the digital twin, requiring many hundreds of hours of compute time on 512 NVIDIA A100 GPUs~\cite{henneking2025bell}.
In this work, the authors presented several optimization steps that led to strong performance gains in the finite element kernels (compared to the baseline performance) on state-of-the-art chips, including AMD's MI300A and NVIDIA's GH200; these optimizations included utilizing GPU shared memory and specifying explicit launch bounds. Even with these optimizations, the total compute time was still largely dominated by the finite element kernels.

The action of a finite element discretized PDE operator on a vector is typically a sparse operation with a FLOP count and memory access rate proportional to the number of DOF; for PDE-based applications, the finite element kernels performing this computation are therefore often the performance bottleneck of the entire application code.
In other words, the highest potential for improving application performance is to increase the kernel throughput (which is the metric-of-interest that determines time-to-solution).
In this paper, we demonstrate that kernel performance of high-order finite element discretizations can be significantly accelerated with the FP64 tensor cores.
We focus on the performance improvement of a particular application example---the tsunami early warning digital twin.
By using the application code and discretized operator presented in \cite{henneking2025bell}, we ensure that the application example, i.e., the particular PDE operator and discretization method, is application-driven rather than specifically chosen to enable the highest performance gains.
Indeed, larger performance gains than the ones presented here could likely be achieved if the operators were discretized in such a way to best fit the tensor-core architecture, as will be discussed in Section~\ref{subsec:performance-limiting-factors}.

\vspace{1em}
The tsunami dynamics are modeled by the acoustic--gravity wave equations that couple the propagation of ocean acoustic waves with surface gravity waves.
This PDE model is derived by linearization of the conservation of mass and momentum around hydrostatic pressure in the compressible ocean, and the coupling to the surface gravity wave comes from a modified free surface boundary condition at the sea surface~\cite{lotto2015tsunami}.
The model takes the form of a coupled first-order system in the velocity vector field $\vec{u}(\vec{x},t)$, the pressure field $p(\vec{x},t)$, and the surface gravity wave height $\eta(\vec{x},t)$,

\begin{equation}
  \left\{
    \begin{aligned}
      \rho\, \p_t \vec{u} + \nabla p &= 0, & \Omega \times (0,T), \\
      K^{-1} \p_t p + \nabla \cdot \vec{u} &= 0, & \Omega \times (0,T), \\
      p &= \rho g \eta, & \p \Omega_{\text s} \times (0,T), \\
      \p_t \eta &= \vec{u} \cdot \vec{n}, & \p \Omega_{\text s} \times (0,T), \\
      \vec{u} \cdot \vec{n} &= -\p_t b, & \p \Omega_{\text b} \times (0,T), \\
      \vec{u} \cdot \vec{n} &= Z^{-1} p, & \p \Omega_{\text a} \times (0,T), \\
    \end{aligned}
    \right.\label{eq:fwd-pde}
  \end{equation}
  with homogeneous initial conditions. Here, $K$ and $\rho$ are the bulk modulus and density of seawater, $Z = \rho c$ is the acoustic wave impedance, $c = \sqrt{K/\rho}$ is the speed of sound in seawater, $g$ is gravitational acceleration, and $\p_t b$ is the seafloor inward normal velocity with $b$ the displacement. The spatial domain is denoted by $\Omega$ with boundaries $\p \Omega_{\text s}$ (sea surface), $\p \Omega_{\text b}$ (sea bottom), and $\p \Omega_{\text a}$ (lateral, absorbing boundaries); $\vec{n}$ is the outward unit normal; and the temporal domain is $(0,T)$.

  The acoustic--gravity model is cast into a mixed variational formulation and then discretized with MFEM using the Galerkin finite element method and explicit fourth-order Runge--Kutta (RK4) time-stepping.

  The finite element discretization uses fourth-order continuous ($H^1$-conforming) scalar-valued pressure and third-order discontinuous ($L^2$-conforming) velocity components.
  The computational expense of the tsunami digital twin application is overwhelmingly dominated by the cost of the acoustic--gravity wave propagation solver.
  In particular, the dominant cost is the repetitive application (four applications per RK4 timestep) of the time-stepping operator
  \bes
  \left[
    \vb u_\delta \ \vb p_\delta
  \right]^T
  = \vb M^{-1}
  \left(
    - \vb A
    \left[
      \vb u \ \vb p
    \right]_i^T
    +
    \left[
      \vb f \ \vb g
    \right]_i^T
  \right) ,
  \ees
  where $\left[ \vb u \ \vb p \right]_i^T$ and $\left[ \vb f \ \vb g \right]_i^T$ are respectively the state and right-hand-side vectors at time instance $i$, and $\left[ \vb u_\delta \ \vb p_\delta \right]$ is the state increment used by RK4. The (lumped) mass matrix $\vb M$ and stiffness matrix $\vb A$ are discretizations of the block operators $M$ and $A$ defined by:
  \bes
  \! \! \! \! \left( \!
    M \!
    \left[
      \begin{array}{@{}c@{}}
        \vec u \\
        p
      \end{array}
    \right] \! , \!
    \left[
      \begin{array}{@{}c@{}}
        \vec \tau \\
        v
      \end{array}
    \right]
  \right)
  \coloneq
  \left[
    \begin{array}{@{}cc@{}}
      (\rho \vec u, \vec \tau) & 0 \\
      0 & \! \! \! (K^{-1} p, v) + \lb(\rho g)^{-1} p, v \rb_{\p \Omega_{\text s}} \\
    \end{array}
  \right]
  \! \!
  \ees
  and
  \be
  \left( \!
    A \!
    \left[
      \begin{array}{@{}c@{}}
        \vec u \\
        p
      \end{array}
    \right] \! , \!
    \left[
      \begin{array}{@{}c@{}}
        \vec \tau \\
        v
      \end{array}
    \right]
  \right)
  \coloneq
  \left[
    \begin{array}{@{}cc@{}}
      0 & (\nabla p, \vec \tau) \\
      -(\vec u, \nabla v) & \lb Z^{-1} p, v \rb_{\p \Omega_{\text a}} \\
    \end{array}
  \right] ,
  \label{eq:block-mult}
  \ee
  where $\vec u, \vec \tau \in (L^2(\Omega))^3$ and $p, v \in H^1(\Omega)$; $(\cdot, \cdot)$ denotes the (componentwise) $L^2(\Omega)$ inner product, and $\lb \cdot, \cdot \rb_{\p \Omega}$ is the $L^2(\p \Omega)$ inner product over (part of) the boundary $\p \Omega$.

  The finite element kernels implementing the action of the operator (\ref{eq:block-mult}) are the main computational expense of solving the PDE model (\ref{eq:fwd-pde}).
  While the runtime spent in these kernels varies by scale and workload, they typically take more than half of the total runtime and in some cases can consume over 90\%, motivating the extensive performance analysis and GPU kernel optimizations presented in Section~\ref{sec:optimizations}.

  \subsection{CUDA and tensor cores}
  On NVIDIA GPUs, CUDA C++ employs a unique architecture called \textit{Single Instruction, Multiple Threads} (SIMT) \cite{simt-arch}.
  The same instruction is executed in groups of 32 threads (called a \textit{warp}), and multiple warps of work are scheduled and pipelined on GPUs to achieve high compute throughput and memory bandwidth.
  The GPU hardware that executes the SIMT instructions is typically called \textit{CUDA cores}.
  In a CUDA C++ kernel, multiple warps of threads are grouped into a thread block, also known as a cooperative thread array.
  Threads in the same thread block communicate via an on-chip, high-bandwidth, low-latency memory called \textit{shared memory}~\cite{shared-memory}.

  Shared memory is divided into 32 \textit{banks}. On current NVIDIA GPUs, the shared memory address space is divided into banks of 4~bytes, i.e.,
  \begin{equation*}
    [\text{bank}] = [\text{address}] / 4\text{ bytes} \!\!\!\mod 32.
  \end{equation*}
  For threads from the same warp, accesses to shared memory addresses in the same bank are serialized.
  This is called a \textit{shared memory bank conflict}, and leads to higher latency and lower throughput for the shared memory load and store instructions.

  In addition to the CUDA cores, the latest NVIDIA GPUs also have \textit{tensor cores} to speed up MMA instructions.
  Later in this work, we use the syntax $m$/$n$/$k$ for matrix-matrix multiplications of shape $m$-by-$k$ times $k$-by-$n$ equals $m$-by-$n$.
  In this work, we use the FP64 \textit{DMMA} instruction that is first available in NVIDIA's Ampere architecture.

%% file: 4_optimizations.tex
In this section, we provide a detailed description of the performance analysis and optimization we performed for the finite element kernels from Section~\ref{sec:background}.
In Section~\ref{subsec:cuda-core-kernels}, we start with the original CUDA core kernels and their performance-limiting factors determined by Nsight Compute profiles.
In Section~\ref{subsec:using-tensor-cores}, we show how to use the FP64 tensor cores by invoking the DMMA instructions.
Next, we describe the techniques we used to avoid shared memory bank conflicts in the FP64 tensor core computations in Section~\ref{subsec:mapping} and Section~\ref{subsec:index-order}.
In Section~\ref{subsec:performance-limiting-factors}, we examine the performance-limiting factors of the DMMA kernels.
In Section~\ref{subsec:choice-of-ptx}, we explain why the CUDA PTX DMMA instruction is used over other frameworks,
and finally discuss the fused PA and fused MF kernel optimizations in Section~\ref{subsec:pa-mf-opts}.

\subsection{The CUDA core kernels}
\label{subsec:cuda-core-kernels}
A sequence of matrix-matrix multiplications is performed as part of the CUDA kernels that implement the actions of finite element discretized PDE operators on a vector (\ref{eq:tensor-notation}).
The kernels used in the Gordon Bell Prize-winning work~\cite{henneking2025bell} use the CUDA cores to perform these matrix-matrix multiplications.
In these CUDA core kernels, for each matrix-matrix multiplication, each thread in the thread block calculates one element of output matrix $C$.
For each thread, the needed elements from input matrices $A$ and $B$ are loaded from shared memory, and the $C$ matrix elements are written back to shared memory.
This results in the following skeleton CUDA code for such matrix-matrix multiplication:

\begin{lstlisting}[language=C++]
int m = threadIdx.x, n = threadIdx.y;
double accC = 0.0;
for (int kk = 0; kk < k; kk++){
  accC += smemA[m, kk] * smemB[kk, n];
}
smemC[m, n] = accC;
\end{lstlisting}
The above CUDA core code leads to a low FLOP-over-byte ratio. For example, a typical matrix-matrix multiplication in the MFEM workflow has $m = 25$, $n = 5$ and $k = 4$.
The total number of bytes loaded and stored to shared memory for the operations in the above code is
\[
  \begin{split}
    &\texttt{sizeof(double)} \times (m \times n \times (k\times2) + m \times n) \\
    &=8\text{ bytes} \times (25 \times 5 \times (4\times2) + 25 \times 5) \\
    &= 9000 \text{ bytes}.
  \end{split}
\]
Note that the same input matrix elements in $A$ and $B$ are loaded by multiple threads, and there is no data sharing among the different threads after the shared memory loads.
The total number of FP64 operations performed is
\[
  m \times n \times k \times 2 = 25 \times 5 \times 4 \times 2 = 1000,
\]
and the FLOP-over-byte ratio is
\[
  1000/9000 \sim 0.11 \text{ [FLOP/byte].}
\]
Therefore, it is the shared memory bandwidth, not the FP64 compute FLOP/s, that limits the performance of the kernel.
This is confirmed by the outputs from the Nsight Compute profile shown in Table~\ref{tab:cuda-core-kernel-ncu}:
the metric ``L1: Data Pipe Lsu Wavefronts'' in the memory throughput breakdown hitting a $97\%$ percentage indicates that the kernel utilizes $97\%$ of the shared memory bandwidth.
In contrast, in the same table, the metric ``SM: Pipe Fp64 Cycles Active'' hits only $14\%$.

\begin{table*}[htbp]
  \centering
  \caption{\upshape{Compute and memory throughput breakdown from the Nsight Compute profile of the original PA kernel on GH200. Based on the profiler's analysis, this kernel takes a total of 21,200,278 cycles. The kernel's performance bottleneck is the shared memory bandwidth (``L1: Data Pipe Lsu Wavefronts'') at $97\%$, achieving an FP64 utilization (``SM: Pipe Fp64 Cycles Active'') of $14\%$.}}
  \label{tab:cuda-core-kernel-ncu}
  \begin{tabular}{p{7.45cm}r@{\hspace{2em}}p{7.45cm}r}
    \toprule
    \multicolumn{2}{c}{\textbf{Compute Throughput Breakdown}} & \multicolumn{2}{c}{\textbf{Memory Throughput Breakdown}} \\
    \midrule
    \rowcolor{GREY!5}
    SM: Inst Executed Pipe Lsu [\%] & 48.49 & \textbf{L1: Data Pipe Lsu Wavefronts [\%]} & \textbf{97.03} \\
    SM: Issue Active [\%] & 45.98 & L2: Lsu Writeback Active [\%] & 73.98 \\
    \rowcolor{GREY!5}
    SM: Inst Executed [\%] & 45.98 & L1: Lsuin Requests [\%] & 48.49 \\
    SM: Mio Inst Issued [\%] & 27.09 & DRAM: Cycles Active [\%] & 43.40 \\
    \rowcolor{GREY!5}
    SM: Mio2rf Writeback Active [\%] & 18.88 & DRAM: Dram Sectors [\%] & 33.30 \\
    SM: Pipe Fmaheavy Cycles Active [\%] & 15.25 & L2: D Sectors Fill Device [\%] & 28.87 \\
    \rowcolor{GREY!5}
    SM: Pipe Alu Cycles Active [\%] & 14.20 & L2: D Sectors [\%] & 24.59 \\
    \textbf{SM: Pipe Fp64 Cycles Active [\%]} & \textbf{13.98} & L2: T Tag Requests [\%] & 20.46 \\
    \rowcolor{GREY!5}
    SM: Pipe Shared Cycles Active [\%] & 13.98 & GPU: Compute Memory Access Throughput Internal Activity [\%] & 20.46 \\
    SM: Mio Pq Read Cycles Active [\%] & 13.19 & L1: Data Bank Reads [\%] & 18.24 \\
    \rowcolor{GREY!5}
    SM: Mio Pq Write Cycles Active [\%] & 13.19 & L2: T Sectors [\%] & 16.86 \\
    SM: Inst Executed Pipe Uniform [\%] & 11.77 & L2: Lts2xbar Cycles Active [\%] & 14.53 \\
    \rowcolor{GREY!5}
    SM: Inst Executed Pipe Adu [\%] & 11.39 & L2: Xbar2lts Cycles Active [\%] & 14.53 \\
    SM: Inst Executed Pipe Cbu Pred On Any [\%] & 8.49 & L1: M L1tex2xbar Req Cycles Active [\%] & 8.75 \\
    \rowcolor{GREY!5}
    SM: Pipe Fma Cycles Active [\%] & 7.63 & L1: Data Bank Writes [\%] & 8.63 \\
    IDC: Request Cycles Active [\%] & 3.00 & L1: M Xbar2l1tex Read Sectors [\%] & 5.89 \\
    \rowcolor{GREY!5}
    SM: Memory Throughput Internal Activity [\%] & 0 & L1: Texin Sm2tex Req Cycles Active [\%] & 0.00 \\
    SM: Instruction Throughput Internal Activity [\%] & 0 & L2: D Sectors Fill Sysmem [\%] & 0.00 \\
    \rowcolor{GREY!5}
    SM: Inst Executed Pipe Xu [\%] & 0 & L2: D Atomic Input Cycles Active [\%] & 0 \\
    SM: Inst Executed Pipe Tex [\%] & 0 & L1: M Xbar2l1tex Read Sectors Mem Dshared [\%] & 0 \\
    \rowcolor{GREY!5}
    SM: Inst Executed Pipe Ipa [\%] & 0 & L1: Tmain Requests [\%] & 0 \\
    SM: Pipe Tensor Cycles Active [\%] & 0 & L1: Data Pipe Tex Wavefronts [\%] & 0 \\
    \rowcolor{GREY!5}
    SM: Pipe Tensor Op Dmma Cycles Active [\%] & 0 & L1: Tex Writeback Active [\%] & 0 \\
    SM: Pipe Tensor Type Hmma Hgmma Qgmma Imma Igmma Brmma Bgmma Cycles Active [\%] & 0 & GPU: Compute Memory Request Throughput Internal Activity [\%] & 0 \\
    \rowcolor{GREY!5}
    SM: Pipe Tma Cycles Active [\%] & 0 & L1: F Wavefronts [\%] & 0 \\
    \bottomrule
  \end{tabular}

  \vspace{3em}

  \caption{\upshape{Compute and memory throughput breakdown from the Nsight Compute profile of the DMMA PA kernel on GH200. Based on the profiler's analysis, this kernel takes a total of 13,871,896 cycles. As for the original kernel (Table~\ref{tab:cuda-core-kernel-ncu}), this kernel's performance bottleneck is still the shared memory bandwidth (``L1: Data Pipe Lsu Wavefronts'')---now at $84\%$---but the DMMA utilization (``Pipe Tensor Op Dmma Cycles Active'') has increased to $54\%$.}}
  \label{tab:tensor-core-kernel-ncu}
  \begin{tabular}{p{7.45cm}r@{\hspace{2em}}p{7.45cm}r}
    \toprule
    \multicolumn{2}{c}{\textbf{Compute Throughput Breakdown}} & \multicolumn{2}{c}{\textbf{Memory Throughput Breakdown}} \\
    \midrule
    \rowcolor{GREY!5}
    SM: Issue Active [\%] & 72.73 & \textbf{L1: Data Pipe Lsu Wavefronts [\%]} & \textbf{84.43} \\
    SM: Inst Executed [\%] & 72.73 & DRAM: Cycles Active [\%] & 66.24 \\
    \rowcolor{GREY!5}
    SM: Pipe Shared Cycles Active [\%] & 57.06 & DRAM: Dram Sectors [\%] & 50.84 \\
    \textbf{SM: Pipe Tensor Op Dmma Cycles Active [\%]} & \textbf{54.08} & L2: D Sectors Fill Device [\%] & 44.07 \\
    \rowcolor{GREY!5}
    SM: Pipe Tensor Cycles Active [\%] & 54.08 & L2: Lsu Writeback Active [\%] & 43.55 \\
    SM: Pipe Alu Cycles Active [\%] & 47.73 & L1: Lsuin Requests [\%] & 41.14 \\
    \rowcolor{GREY!5}
    SM: Inst Executed Pipe Lsu [\%] & 41.14 & L2: D Sectors [\%] & 37.42 \\
    SM: Pipe Fmaheavy Cycles Active [\%] & 38.04 & L2: T Tag Requests [\%] & 32.84 \\
    \rowcolor{GREY!5}
    SM: Mio Pq Write Cycles Active [\%] & 25.21 & GPU: Compute Memory Access Throughput Internal Activity [\%] & 32.82 \\
    SM: Mio Pq Read Cycles Active [\%] & 24.87 & L2: T Sectors [\%] & 25.58 \\
    \rowcolor{GREY!5}
    SM: Mio Inst Issued [\%] & 24.29 & L2: Lts2xbar Cycles Active [\%] & 22.20 \\
    SM: Pipe Fma Cycles Active [\%] & 19.02 & L2: Xbar2lts Cycles Active [\%] & 21.42 \\
    \rowcolor{GREY!5}
    SM: Inst Executed Pipe Adu [\%] & 14.90 & L1: Data Bank Reads [\%] & 15.52 \\
    SM: Inst Executed Pipe Uniform [\%] & 14.84 & L1: Data Bank Writes [\%] & 12.75 \\
    \rowcolor{GREY!5}
    SM: Mio2rf Writeback Active [\%] & 11.29 & L1: M L1tex2xbar Req Cycles Active [\%] & 12.72 \\
    SM: Inst Executed Pipe Cbu Pred On Any [\%] & 8.07 & L1: M Xbar2l1tex Read Sectors [\%] & 9.02 \\
    \rowcolor{GREY!5}
    IDC: Request Cycles Active [\%] & 3.21 & L1: Texin Sm2tex Req Cycles Active [\%] & 0.00 \\
    SM: Pipe Fp64 Cycles Active [\%] & 2.98 & L2: D Sectors Fill Sysmem [\%] & 0.00 \\
    \rowcolor{GREY!5}
    SM: Instruction Throughput Internal Activity [\%] & 0 & L2: D Atomic Input Cycles Active [\%] & 0 \\
    SM: Memory Throughput Internal Activity [\%] & 0 & L1: Data Pipe Tex Wavefronts [\%] & 0 \\
    \rowcolor{GREY!5}
    SM: Inst Executed Pipe Xu [\%] & 0 & L1: Tmain Requests [\%] & 0 \\
    SM: Inst Executed Pipe Tex [\%] & 0 & L1: Tex Writeback Active [\%] & 0 \\
    \rowcolor{GREY!5}
    SM: Inst Executed Pipe Ipa [\%] & 0 & L1: M Xbar2l1tex Read Sectors Mem Dshared [\%] & 0 \\
    SM: Pipe Tensor Type Hmma Hgmma Qgmma Imma Igmma Brmma Bgmma Cycles Active [\%] & 0 & GPU: Compute Memory Request Throughput Internal Activity [\%] & 0 \\
    \rowcolor{GREY!5}
    SM: Pipe Tma Cycles Active [\%] & 0 & L1: F Wavefronts [\%] & 0 \\
    \bottomrule
  \end{tabular}
\end{table*}

\subsection{Using tensor cores}
\label{subsec:using-tensor-cores}
By using tensor cores instead of CUDA cores for FP64 matrix-matrix multiplications, with the same shapes, fewer bytes are loaded from the shared memory.
\emph{As a result, using the DMMA instruction directly alleviates the bottleneck on shared memory bandwidth}.

With the DMMA instruction, each warp in the thread block performs matrix-matrix multiplications collectively.
Each thread of the warp loads a unique set of matrix elements from input matrices $A$ and $B$: each matrix element in $A$ and $B$ is loaded only once among the threads in a warp.
The $m8n8k4$ DMMA instruction works in the following way~\cite{cuda-programming-guide-mma-fragments}:
\begin{enumerate}
  \item Each thread in the warp loads one FP64 element of matrix A ($a0$) from shared memory: the map between the logical row and column indices and the lane index of the warp are shown in Table~\ref{tab:mma-a-frag}.
  \item Each thread in the warp loads one FP64 element of matrix B ($b0$) from shared memory: the map between the logical row and column indices and the lane index of the warp are shown in Table~\ref{tab:mma-b-frag}.
  \item Invoke the CUDA PTX instruction for DMMA with $a0$ and $b0$ from each thread as input~\cite{cuda-programming-guide-mma-instructions}. The CUDA PTX instruction can be seamlessly inserted into the CUDA C++ code with the inline PTX syntax.
  \item The DMMA instruction returns the result matrix C:\@ each thread in a warp gets two FP64 elements ($c0$ and $c1$) for a total of $64$ ($=32 \times 2$) elements. The maps between the logical row and column indices and the lane index of the warp are shown in Table~\ref{tab:mma-c-frag}.
\end{enumerate}
This results in the following skeleton CUDA code for performing matrix-matrix multiplications with DMMA tensor cores with the CUDA PTX DMMA instruction:
\begin{lstlisting}[language=C++]
int lane_idx = threadIdx.x %
double a[1], b[1], c[2];
a[0] = smemA[... lane_idx...];  // load a0
b[0] = smemB[... lane_idx...];  // load b0
asm volatile(
    "mma.sync.aligned.m8n8k4.row.col.f64.f64."
    "f64.f64 {%
    : "+d"(c[0]), "+d"(c[1])
    : "d"(a[0]), "d"(b[0]));    // invoke DMMA
smemC[... lane_idx...] = c[0];  // store c0
smemC[... lane_idx...] = c[1];  // store c1
\end{lstlisting}
The total number of bytes loaded from and stored to shared memory using tensor cores now becomes %
\[
  \begin{split}
    &\texttt{sizeof(double)} \times (m \times k + n \times k + m \times n) \\
    &=8\text{ bytes} \times (25 \times 4 + 5 \times 4 + 25 \times 5) \\
    &=1960\text{ bytes}.
  \end{split}
\]
Fewer bytes are loaded from shared memory versus the CUDA core implementation, because the input A/B matrix values are shared among the threads invoking the DMMA instruction in the same warp and only need to be loaded once per warp.
To perform the target matrix-matrix multiplication with the DMMA instruction, the logical $m$/$n$/$k$ indices of the DMMA instruction ($m_i$/$n_i$/$k_i$) are mapped to the target problem's $m$/$n$/$k$ indices ($m_p$/$n_p$/$k_p$):
\begin{align*}
  f_m &: m_i \mapsto m_p , \\
  f_n &: n_i \mapsto n_p , \\
  f_k &: k_i \mapsto k_p .
\end{align*}

\begin{table*}[htb]
  \centering
  \begin{minipage}{0.45\linewidth}
    \centering
    \caption{\upshape{The map between the logical row and column indices and the lane index of a warp for matrix A of the $m8n8k4$ DMMA instruction, shown as \texttt{T[lane number]:a0}.
    Quoted from \cite{cuda-programming-guide-mma-fragments}.}}
    \label{tab:mma-a-frag}
    \vspace{-0.75em}
    \begin{tabular}{ccccc}
      \toprule
      Row/Col & \texttt{0} & \texttt{1} & \texttt{2} & \texttt{3} \\
      \midrule
      \rowcolor{GREY!5}
      \texttt{0} & \texttt{T0}:\texttt{a0} & \texttt{T1}:\texttt{a0} & \texttt{T2}:\texttt{a0} & \texttt{T3}:\texttt{a0} \\
      \texttt{1} & \texttt{T4}:\texttt{a0} & \texttt{T5}:\texttt{a0} & \texttt{T6}:\texttt{a0} & \texttt{T7}:\texttt{a0} \\
      \rowcolor{GREY!5}
      \texttt{...} & \texttt{...} & \texttt{...} & \texttt{...} & \texttt{...} \\
      \texttt{7} & \texttt{T28}:\texttt{a0} & \texttt{T29}:\texttt{a0} & \texttt{T30}:\texttt{a0} & \texttt{T31}:\texttt{a0} \\
      \bottomrule
    \end{tabular}
  \end{minipage}
  \begin{minipage}{0.45\linewidth}
    \centering
    \caption{\upshape{The map between the logical row and column indices and the lane index of a warp for matrix B of the $m8n8k4$ DMMA instruction, shown as \texttt{T[lane number]:b0}.
    Quoted from \cite{cuda-programming-guide-mma-fragments}.}}
    \label{tab:mma-b-frag}
    \vspace{-0.75em}
    \begin{tabular}{ccccc}
      \toprule
      Row/Col & \texttt{0} & \texttt{1} & \texttt{...} & \texttt{7} \\
      \midrule
      \rowcolor{GREY!5}
      \texttt{0} & \texttt{T0}:\texttt{b0} & \texttt{T4}:\texttt{b0} & \texttt{...} & \texttt{T28}:\texttt{b0} \\
      \texttt{1} & \texttt{T1}:\texttt{b0} & \texttt{T5}:\texttt{b0} & \texttt{...} & \texttt{T29}:\texttt{b0} \\
      \rowcolor{GREY!5}
      \texttt{2} & \texttt{T2}:\texttt{b0} & \texttt{T6}:\texttt{b0} & \texttt{...} & \texttt{T30}:\texttt{b0} \\
      \texttt{3} & \texttt{T3}:\texttt{b0} & \texttt{T7}:\texttt{b0} & \texttt{...} & \texttt{T31}:\texttt{b0} \\
      \bottomrule
    \end{tabular}
  \end{minipage}

  \vspace{2.0em}

  \caption{\upshape{The map between the logical row and column indices and the lane index of a warp for matrix C\\
  of the $m8n8k4$ DMMA instruction, shown as \texttt{T[lane number]:c[0,1]}. Quoted from \cite{cuda-programming-guide-mma-fragments}.}}
  \label{tab:mma-c-frag}
  \vspace{-0.75em}
  \begin{tabular}{cccccccc}
    \toprule
    Row/Col & \texttt{0} & \texttt{1} & \texttt{2} & \texttt{3} & \texttt{...} & \texttt{6} & \texttt{7} \\
    \midrule
    \rowcolor{GREY!5}
    \texttt{0} & \texttt{T0}:\texttt{c0} & \texttt{T0}:\texttt{c1}  & \texttt{T1}:\texttt{c0} & \texttt{T1}:\texttt{c1}  & \texttt{...}  & \texttt{T3}:\texttt{c0} & \texttt{T3}:\texttt{c1} \\
    \texttt{1} & \texttt{T4}:\texttt{c0} & \texttt{T4}:\texttt{c1}  & \texttt{T5}:\texttt{c0} & \texttt{T5}:\texttt{c1}  & \texttt{...}  & \texttt{T7}:\texttt{c0} & \texttt{T7}:\texttt{c1} \\
    \rowcolor{GREY!5}
    \texttt{2} & \texttt{T8}:\texttt{c0} & \texttt{T8}:\texttt{c1}  & \texttt{T9}:\texttt{c0} & \texttt{T9}:\texttt{c1}  & \texttt{...} & \texttt{T11}:\texttt{c0} & \texttt{T11}:\texttt{c1} \\
    \texttt{3} & \texttt{T12}:\texttt{c0} & \texttt{T12}:\texttt{c1} & \texttt{T13}:\texttt{c0} & \texttt{T13}:\texttt{c1} & \texttt{...} & \texttt{T15}:\texttt{c0} & \texttt{T15}:\texttt{c1} \\
    \rowcolor{GREY!5}
    \texttt{4} & \texttt{T16}:\texttt{c0} & \texttt{T16}:\texttt{c1} & \texttt{T17}:\texttt{c0} & \texttt{T17}:\texttt{c1} & \texttt{...} & \texttt{T19}:\texttt{c0} & \texttt{T19}:\texttt{c1} \\
    \texttt{5} & \texttt{T20}:\texttt{c0} & \texttt{T20}:\texttt{c1} & \texttt{T21}:\texttt{c0} & \texttt{T21}:\texttt{c1} & \texttt{...} & \texttt{T23}:\texttt{c0} & \texttt{T23}:\texttt{c1} \\
    \rowcolor{GREY!5}
    \texttt{6} & \texttt{T24}:\texttt{c0} & \texttt{T24}:\texttt{c1} & \texttt{T25}:\texttt{c0} & \texttt{T25}:\texttt{c1} & \texttt{...} & \texttt{T27}:\texttt{c0} & \texttt{T27}:\texttt{c1} \\
    \texttt{7} & \texttt{T28}:\texttt{c0} & \texttt{T28}:\texttt{c1} & \texttt{T29}:\texttt{c0} & \texttt{T29}:\texttt{c1} & \texttt{...} & \texttt{T31}:\texttt{c0} & \texttt{T31}:\texttt{c1} \\
    \bottomrule
  \end{tabular}
  \vspace{-1.0em}
\end{table*}

For example, in order to perform an $m = 25$, $n = 5$, $k = 4$ matrix-matrix multiplication, we can use four warps for the DMMA instruction with the following maps:
\begin{align*}
  f_m &: m_i \mapsto m_p,\ m_p = m_i + w \times 8,\\
  f_n &: n_i \mapsto n_p,\ n_p = n_i,\\
  f_k &: k_i \mapsto k_p,\ k_p = k_i,
\end{align*}
where $w$ is the warp index, and $f_n$ and $f_k$ are identity maps. $f_m$ can also be the following one:
\[ f_m: m_i \mapsto m_p,\ m_p = m_i \times 4 + w. \]

\subsection{Optimal mapping for bank conflict avoidance}
\label{subsec:mapping}
For different shapes of matrix-matrix multiplications, different $f_m$/$f_n$/$f_k$ can be chosen to avoid shared memory bank conflicts.
For example, in order to perform an $m = 25$, $n=5$, $k=4$ matrix-matrix multiplication, we use four warps for the DMMA instruction with the following maps:
\begin{equation}
  \begin{split}
    f_m &: m_i \mapsto m_p,\ m_p = m_i + w \times 8, \\
    f_n &: n_i \mapsto n_p, \\
    & \ \ [0,1,2,3,4,5,6,7] \rightarrow [0,2,1,3,4,5,6,7], \\
    f_k &: k_i \mapsto k_p,\ k_p = k_i. \label{eq:fm-fn-fk}
  \end{split}
\end{equation}
When each thread (also referred to as a lane in a warp) in a warp loads (stores) an 8-byte word from (to) the shared memory, the load (store) is done in two accessing phases.
In the first phase, lanes 0--15 load (store) from (to) shared memory with each lane occupying two shared memory banks; in the second phase, lanes 16--31 do the same.
As a result, if the first or the second half of the warp accesses 16 distinct shared memory banks during one 8-byte shared memory access, there is no shared memory bank conflict.
With the maps shown in (\ref{eq:fm-fn-fk}), each lane of a warp accesses one 8-byte FP64 number for \texttt{a0}, \texttt{b0}, \texttt{c0} and \texttt{c1}.

Figure~\ref{fig:smem-banks} depicts the shared memory banks being accessed during the two access phases for matrices A, B and C.
There are no shared memory bank conflicts for any of these accesses.
Apart from the $m25n5k4$ ($m=25, n=5, k=4$) matrix shape shown above, we have also tuned the $f_m$/$f_n$/$f_k$ for the following matrix shapes such that they are free from shared memory bank conflicts: $m25n5k5$, $m25n4k5$, $m20n4k5$, $m16n4k5$, $m16n5k4$ and $m20n5k4$.

\subsection{Tensor index reordering for bank conflict avoidance}
\label{subsec:index-order}
The tensor contractions in equation (\ref{eq:tensor-notation}) can be decomposed into three matrix-matrix multiplications, for instance:
\begin{align*}
  X^1_{ajk} & = B^{1d}_{ai} X_{ijk} , \\
  X^2_{abk} & = B^{1d}_{bj} X^1_{ajk} , \\
  Y_{abc}   & = B^{1d}_{ck} X^2_{abk} .
\end{align*}
Here the indices of the tensors also indicate the order in which the numbers are stored in shared memory:
for example, for $X^1_{ajk}$, the numbers are stored in lexicographical order with $a$ being the fastest-changing index,
$j$ being the second-fastest-changing index, and $k$ being the slowest-changing index.
When the index that is being summed over is the second-fastest-changing index (i.e., the index is in the middle), shared memory bank conflicts become inevitable.
This issue can be avoided by using a \textit{cyclic order}, e.g.,
\begin{align*}
  X^1_{jka} & = B^{1d}_{ai} X_{ijk} , \\
  X^2_{kab} & = B^{1d}_{bj} X^1_{jka} , \\
  Y_{abc}   & = B^{1d}_{ck} X^2_{kab} .
\end{align*}
This way the index that is being summed over is always the fastest-changing index.

\subsection{Tensor core kernel performance limiters}
\label{subsec:performance-limiting-factors}
\begin{figure*}[htbp]
  \centering
  \includegraphics[width=0.85\textwidth,trim={0 0 0 0}]{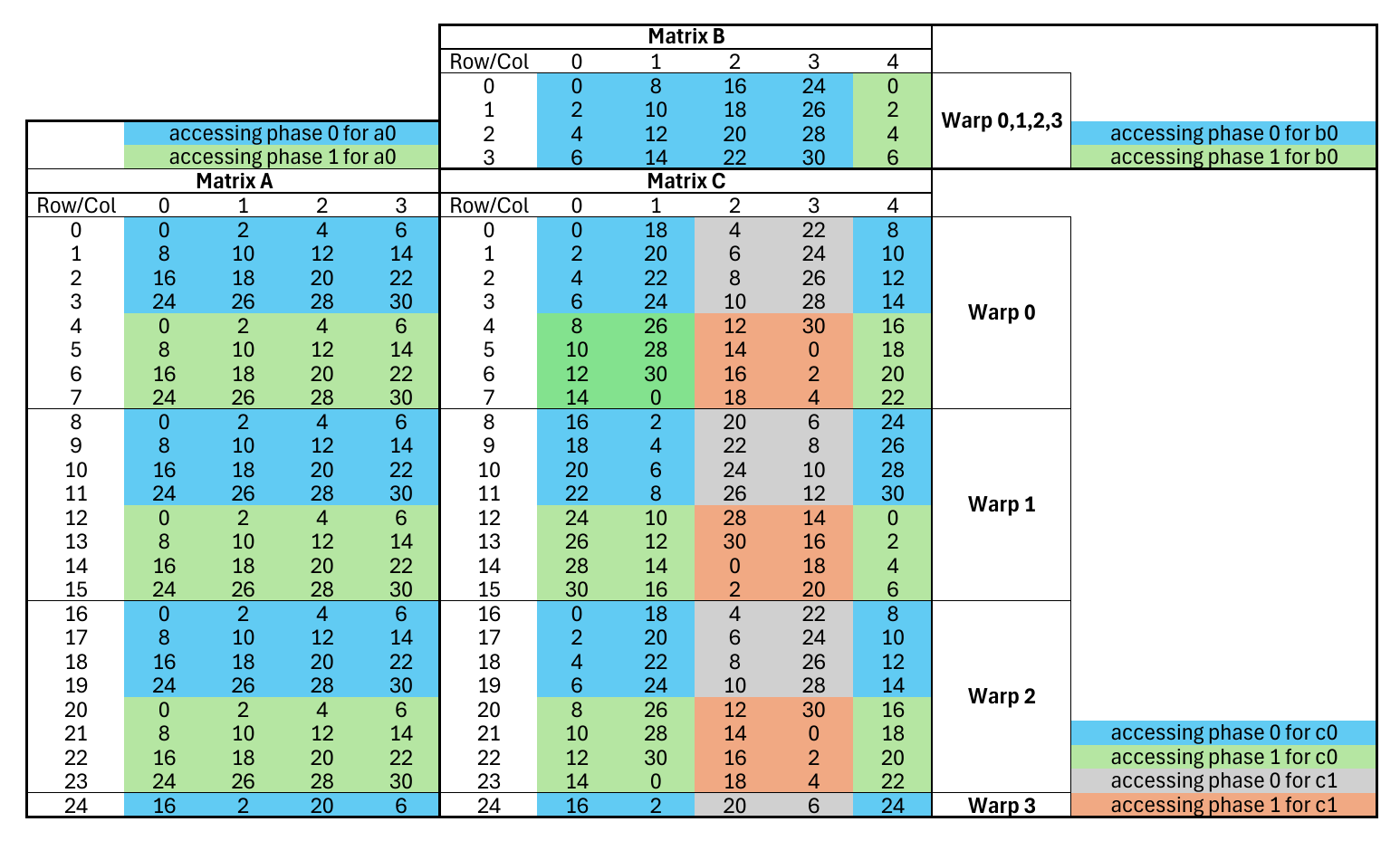}
  \caption{Shared memory banks for matrix A/B/C of an $m = 25$, $n=5$, $k=4$ matrix-matrix multiplication with the maps shown in (\ref{eq:fm-fn-fk}). In each of the phases, the 16 lanes access 16 distinct banks; thus, there are no shared memory bank conflicts.}
  \label{fig:smem-banks}
\end{figure*}

The profile of the DMMA PA kernel in Table~\ref{tab:tensor-core-kernel-ncu} shows that shared memory bandwidth (``L1: Data Pipe Lsu Wavefronts'') is still the highest-used resource in the memory throughput analysis, but the percentage is lower compared to the original PA kernel using CUDA cores; items in the compute throughput breakdown now have a higher percentage, implying the DMMA better utilizes the GPU compute resources.
Specifically, ``SM: Pipe Tensor Op Dmma Cycles Active'' indicates that the DMMA pipe of the GH200 GPU has $54\%$ usage. Overall, the DMMA PA kernel achieves a $1.5\times$ speedup (ratio of the total cycle count for each kernel: $1.5$ = $21{,}200{,}278$ cycles / $13{,}871{,}896$ cycles) over the original PA kernel.

The DMMA PA kernel has a $54\%$ utilization of the DMMA pipe on the GH200, while the original PA kernel has a $14\%$ utilization of the FP64 pipe (``SM: Pipe Fp64 Cycles Active''). One might ask why the $54$ / $14$ = $3.8\times$ increase in compute utilization only leads to $1.5\times$ speedup for the entire kernel: this is primarily because when using the $m8n8k4$ DMMA instruction to perform the various matrix-matrix multiplications, the shapes of these ($m=25$, $n=5$, $k=4$) matrix-matrix multiplications do not exactly fit the ($m=8$, $n=8$, $k=4$) instruction shape, thus wasting a large percentage of the computation due to the mismatch.

\subsection{Choosing CUDA DMMA instruction over other frameworks}
\label{subsec:choice-of-ptx}
In this work, the CUDA PTX DMMA instruction is chosen over other CUDA libraries like CUTLASS~\cite{cutlass} or CUBLAS~\cite{cublas}, because our small $\mathcal{O}(10)$ matrices—such as $25\times5\times4$—require precise control for custom thread-to-fragment mappings (\ref{eq:fm-fn-fk}) and bank conflict elimination (Figure~\ref{fig:smem-banks}).
CUTLASS is optimized for larger GEMMs and lacks the needed granularity; its primary method to avoid shared memory bank conflicts is swizzling,
which requires much larger alignments (e.g., 128 bytes) and shapes (e.g., $m64n64k32$).
While padding could enable the use of CUTLASS kernel-level interfaces, with the small matrix sizes we are targeting, the majority of the shared memory bandwidth and FLOP/s would be wasted on the padding, rendering CUTLASS much less efficient than our approach in this case.

CUBLAS, on the other hand, does not support such unconventional scenarios with small, irregular GEMMs and kernel fusions.
CUBLASDx~\cite{cublasdx}, which is still under development, addresses our needs and may in the future provide an efficient alternative to our approach.

Finally, we note that invoking the CUDA PTX DMMA instruction with the inline PTX syntax, as previously shown in the skeleton CUDA code for DMMA tensor cores,  requires just about the same amount of developer effort as invoking other CUDA device intrinsic functions when writing a CUDA C++ kernel.

\subsection{PA and MF kernel optimizations}
\label{subsec:pa-mf-opts}
\begin{figure*}[htb]
  \centering
  \includegraphics[width=\textwidth,trim={0 0.5in 0 0}]
  {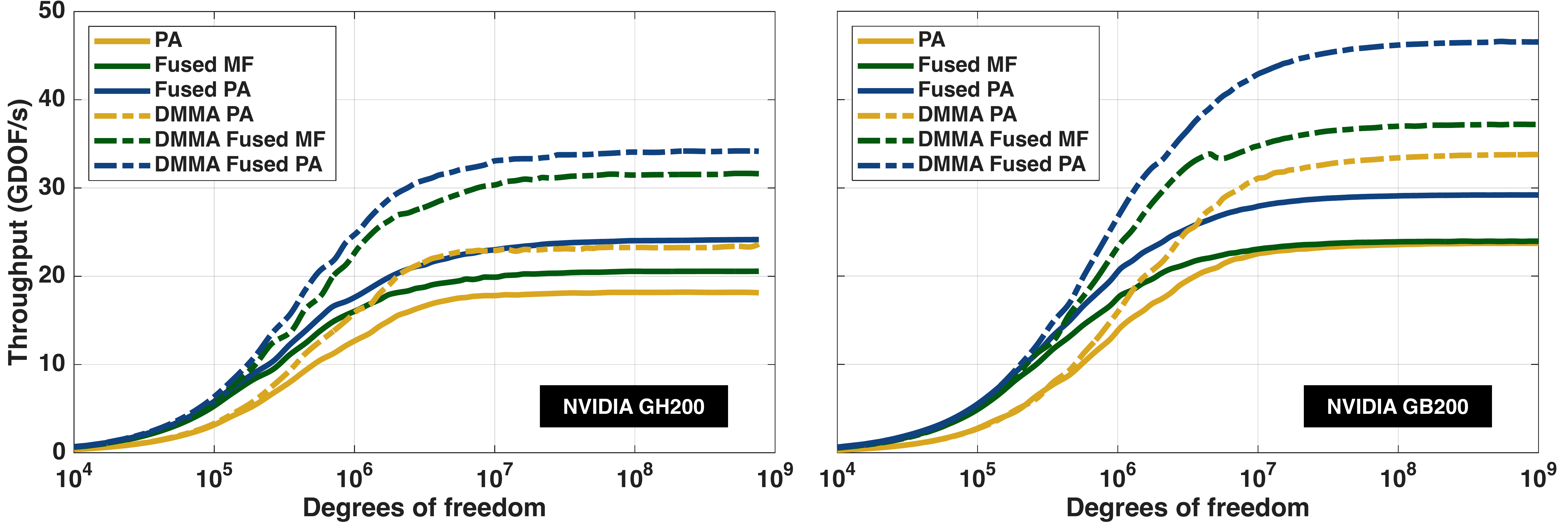}
  \caption{Throughput, in billion degrees of freedom (GDOF) per second, for the finite element kernels corresponding to the off-diagonal blocks in (\ref{eq:block-mult}) on a single NVIDIA GH200 Grace Hopper Superchip (left) and a single NVIDIA GB200 Grace Blackwell Superchip (right). Fusing loops into a single kernel and using tensor-core-enabled DMMA instructions (``DMMA Fused PA'') together yields a 2$\times$ performance gain compared to the original (``PA'') kernel.}
  \label{fig:kernel-performance}
\end{figure*}

Our GPU kernel optimizations focused on the $G^T B^T D B G$ operations in (\ref{eq:mfem-operator}), excluding the application of $P$ and $P^T$, which are sparse matrix-vector products handling MPI communication.
The finite element operator is mixed, i.e., the trial and test spaces are different, and so we need distinct $B_{\text{trial}}$ and $B_{\text{test}}$ operators (and their respective transposes).

By eliminating the initial $G_{\text{test}}$ from the preceding vector operation, the primary kernel $K_{u,p}$ comprises a sequence of two subkernels, $K_u$ and $K_p$,
such that $K_{u,p} = K_p K_u$, where
\[
  K_u = G^T_{\text{trial}} B^T_{\text{trial}} D^T B_{\text{test}}
  \enspace \text{and} \enspace
  K_p = B^T_{\text{test}} D B_{\text{trial}} G_{\text{trial}} .
\]
This can be reordered into three kernels $K_1$, $K_2$, and $K_3$:
\begin{equation*}
  \begin{aligned}
    K_1 &= B^T_{\text{trial}} D^T B_{\text{test}} , \\
    K_2 &= G_{\text{trial}} G^T_{\text{trial}} , \\
    K_3 &= B^T_{\text{test}} D B_{\text{trial}} .
  \end{aligned}
\end{equation*}
Here, $K_2$ is an MFEM operation that can be optimized by introducing indirect reads from a DOF-mapping array and writing back to global memory via atomic operations.
The fused operator $K_{\text{fused}}$ is then expressed as:
\[
  K_{\text{fused}} = B^T_{\text{test}} D B_{\text{trial}} G_{\text{trial}} G^T_{\text{trial}} B^T_{\text{trial}} D^T B_{\text{test}} ,
\]
and further optimized to:
\[
  K_{\text{fused}} = B^T_{\text{test}} B^T_{\text{trial}} D D^T B_{\text{trial}} B_{\text{test}} .
\]

The primary benefit of this loop fusion is a twofold reduction in PA data movement, as the applications of $D$ and $D^T$ occur within the same kernel.
However, even after applying all available memory optimizations---such as avoiding caching of large vectors, recomputing values on-the-fly,
and reusing temporary vectors from the RK4 time-stepping algorithm---PA data still accounts for a substantial portion of the total memory usage.

The fused MF kernel optimizations address this issue by eliminating storage of the PA data at quadrature points,
instead requiring additional input vectors and on-the-fly computations, including DOF mappings, basis function arrays, and mesh coordinates for Jacobian computations.
As the fused kernel expands and demands more input data, minimizing shared memory usage per thread block becomes critical to increase GPU occupancy;
this is achieved by relocating $B_{\text{trial}}$ and $B_{\text{test}}$ data to constant memory and shuffling intermediate computations to reuse temporary shared memory space.

%% file: 5_results.tex
In this section, we evaluate the effect of kernel fusion and tensor core (DMMA) optimizations on the performance of the PA and MF algorithms.
We report single-GPU throughput and performance per Watt for GH200 and GB200, with significant benefits in both metrics.
We also demonstrate excellent weak and strong scalability on up to 9,216 GH200 GPUs.

\subsection{Single-GPU performance}
Figure~\ref{fig:kernel-performance} shows the throughput performance of the finite element kernels on NVIDIA's GH200 Grace Hopper Superchip (left) and GB200 Grace Blackwell Superchip (right). For the various kernel types, the DMMA optimizations achieve 35\% to 59\% speedup over the original non-DMMA kernels in the saturated regime (i.e., for sufficiently many DOFs).
Fusing loops into a single kernel and using tensor-core-enabled DMMA instructions (“DMMA Fused PA”) together yields a 2$\times$ performance gain compared to the original PA kernel.

\begin{table*}[htb]
  \centering
  \caption{Throughput performance and performance per Watt for each kernel on GH200 and GB200 for a problem of $540$ million DOF.}
  \label{tab:perf-per-watt}
  \begin{tabular}{llrrrrrr}
    \toprule
    Metric & Chip & PA & Fused PA & Fused MF & DMMA PA & DMMA Fused PA & DMMA Fused MF \\
    \midrule
    \rowcolor{GREY!5}
    GDOF/s & GB200 & 23.78 & 29.28 & 24.01 & 33.72 & 46.60 & 37.26 \\
    GDOF/s & GH200 & 18.73 & 24.04 & 20.45 & 25.27 & 36.15 & 32.24 \\
    \rowcolor{GREY!5}
    MDOF/Watt & GB200 & 26.60 & 36.41 & 37.01 & 31.37 & 45.70 & 49.03 \\
    MDOF/Watt & GH200 & 28.65 & 40.14 & 41.02 & 36.51 & 52.42 & 50.88 \\
    \rowcolor{GREY!5}
    AVG Power [Watt] & GB200 & 893.84 & 804.07 & 648.79 & 1074.76 & 1019.63 & 759.97 \\
    AVG Power [Watt] & GH200 & 653.72 & 598.88 & 498.58 & 692.19 & 689.64 & 633.69 \\
    \bottomrule
  \end{tabular}
  \vspace{-1.5em}
\end{table*}

Table~\ref{tab:perf-per-watt} details the throughput performance (GDOF/s) and the performance per Watt (MDOF/W)
on both GH200 and GB200 for the original and the DMMA kernels, for a problem of 540 million DOF.
In addition to higher performance, the DMMA kernels also achieve higher performance per Watt than the original kernels.
Usage of FP64 tensor cores improves performance per Watt by 18\% on GB200 and 27\% on GH200.
When fusion is added, performance per Watt improves overall by 72\% on GB200 and 83\% on GH200.
We note that GB200 achieves a lower performance per Watt than GH200 for each of the kernel types.
The reasons for this include: (1) GB200 has higher idle GPU power, (2) GB200 runs at a 4\% higher clock frequency which,
while adding additional performance, can lead to less efficient computations,
and (3) GB200 has much higher memory bandwidth and tensor core performance for lower precisions (FP16, BF16, FP8, FP6 and FP4),
while the kernels of interest here do not make high utilizations of any of these compute resources.

The fused PA implementation achieves a lower percentage of the GPU's theoretical peak FLOP/s compared to the fused MF version but ultimately delivers faster overall runtime.
This counterintuitive outcome arises because, although fused MF exhibits higher arithmetic intensity (performing more computations per byte of data transferred) and transfers less data per DOF, its higher FLOP/DOF ratio increases the overall computational workload.
As a result, the gains in FLOP/s efficiency are outweighed by the additional FLOPs required, leading to longer execution times despite the apparent performance advantage in peak utilization.

\subsection{Multi-GPU performance}
Scalability of the MFEM implementation was tested on CSCS's \emph{Alps} system.
\emph{Alps} is an HPE Cray EX supercomputer with 2,688 nodes, each containing 4 NVIDIA GH200 Superchips.
Each GH200 integrates a Grace CPU (72 Arm cores) and an H100 GPU (96~GB HBM3, 34~TFLOP/s FP64 and 67~TFLOP/s FP64 tensor core peak throughput).
The system is connected by an HPE Slingshot-11 dragonfly interconnect with at most three network hops between any two nodes.
The theoretical system peak is 574.8~PFLOP/s; and the system achieved a peak of 434.9~PFLOP/s securing the \#8 spot in the November 2025 TOP500 list~\cite{top500}.

\begin{table}[htb]
  \centering
  \caption{Scalability setup on Alps}
  \label{tab:scaling-setup}
  \begin{tabular}{rrrrr}
    \toprule
    & & & Weak scaling & Strong scaling \\
    \cmidrule{4-5}
    Nodes & GPUs & Processor grid & Mesh elements & Elements/GPU \\
    \midrule
    \rowcolor{GREY!5}
    36 & 144 & 2 $\times$ \hspace{1pt} 18 $\times$ 4 & 566,231,040 & 3,932,160 \\
    72 & 288 & 4 $\times$ \hspace{1pt} 18 $\times$ 4 & 1,132,462,080 & 1,966,080 \\
    \rowcolor{GREY!5}
    144 & 576 & 4 $\times$ \hspace{1pt} 36 $\times$ 4 & 2,264,924,160 & 983,040 \\
    288 & 1,152 & 8 $\times$ \hspace{1pt} 36 $\times$ 4 & 4,529,848,320 & 491,520 \\
    \rowcolor{GREY!5}
    576 & 2,304 & 8 $\times$ \hspace{1pt} 72 $\times$ 4 & 9,059,696,640 & 245,760 \\
    1,152 & 4,608 & 16 $\times$ \hspace{1pt} 72 $\times$ 4 & 18,119,393,280 & 122,880 \\
    \rowcolor{GREY!5}
    2,304 & 9,216 & 16 $\times$ 144 $\times$ 4 & 36,238,786,560 & 61,440 \\
    \bottomrule
  \end{tabular}
\end{table}

The setup for the scalability study is shown in Table~\ref{tab:scaling-setup}.
The base case for both strong and weak scalability is defined on 36 nodes (144 GH200), and results are reported up to the full system size of 2,304 nodes (9,216 GH200) available at the time of benchmarking---a 64$\times$ increase of nodes from the base case.
The largest (weak scaling) run solved a problem instance with a total of $\sim$9.28 trillion DOF, averaging $\sim$1.01 billion DOF per GH200. The strong scaling problem was defined as the largest problem instance fitting on the 36 node (144 GH200) base configuration, with a total of $\sim$145 billion DOF.

\begin{figure*}[htb]
  \centering
  \includegraphics[width=\textwidth,trim={0 0.8in 0 0.5in}]
  {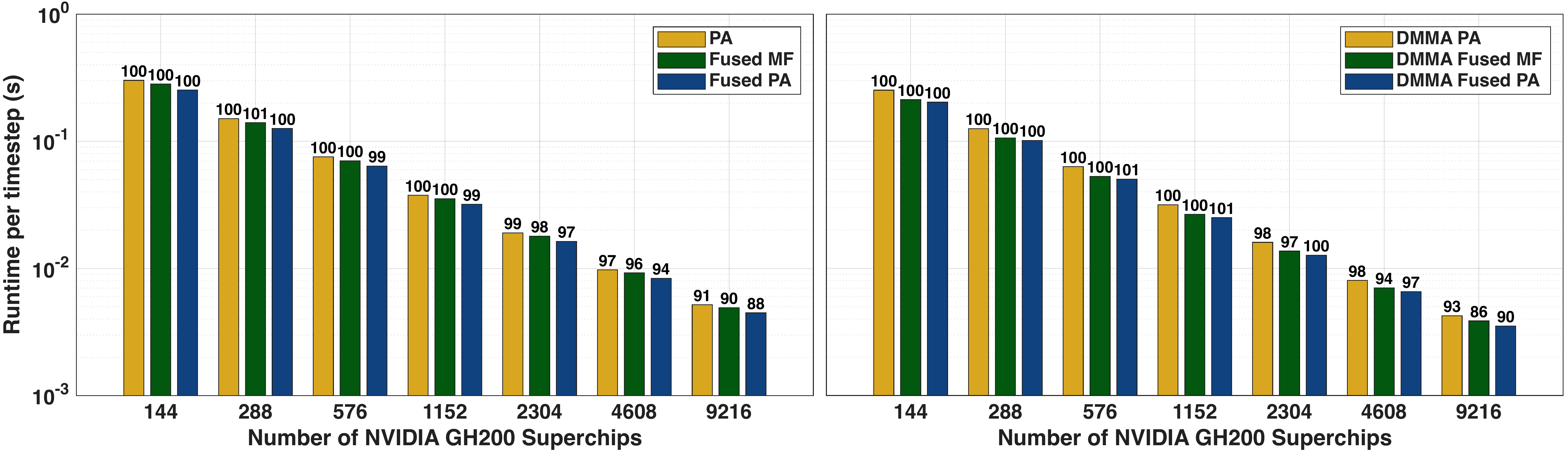}
  \caption{Strong scalability of the MFEM finite element solver on \emph{Alps}, from 36 nodes (144 NVIDIA GH200 Superchips) to 2,304 nodes (9,216 GH200).
    Black numbers above each bar indicate the corresponding parallel efficiencies.
  The three kernel implementations---PA, Fused MF and Fused PA---(left plot) and their respective FP64 tensor-core-accelerated versions (DMMA; right plot) achieve excellent strong scaling (86--91\%) over a 64$\times$ increase of nodes.}
  \label{fig:strong-scaling}
  \vspace{-1.5em}
\end{figure*}

\begin{figure}[htb]
  \centering
  \includegraphics[width=\columnwidth,trim={0 0.3in 0 0}]
  {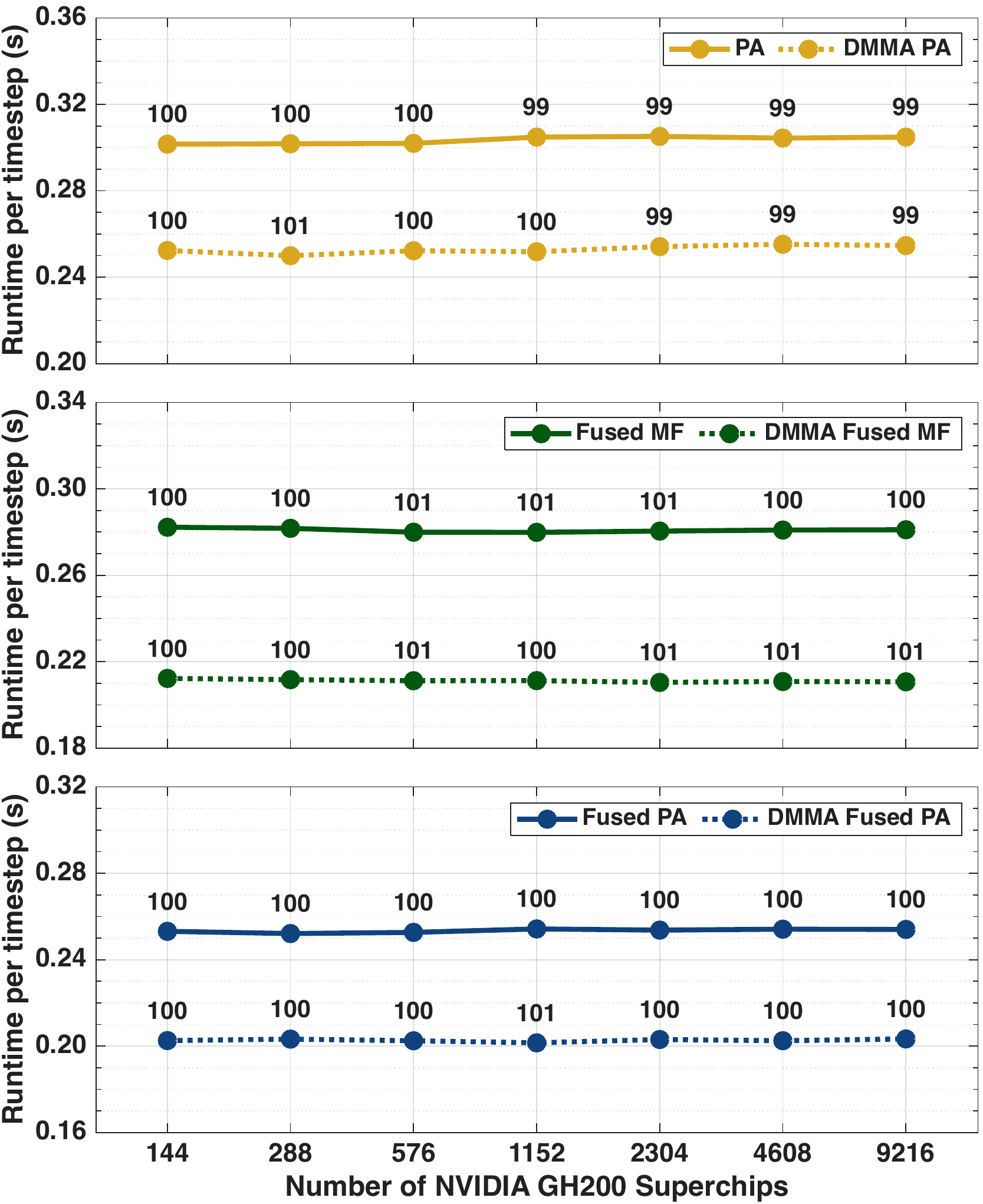}
  \caption{Weak scalability of the MFEM finite element solver on \emph{Alps}, from 36 nodes (144 GH200) to 2,304 nodes (9,216 GH200).
    Numbers along the graph lines indicate parallel efficiency.
  Each of the three kernel implementations---PA (top), Fused MF (middle) and Fused PA (bottom)---and their respective FP64 tensor-core-accelerated versions (DMMA) achieve ideal (linear) weak scaling over a 64$\times$ increase of nodes.}
  \label{fig:weak-scaling}
  \vspace{-1.5em}
\end{figure}

Scalability results are reported in terms of \emph{runtime per timestep}, which is the time-to-solution metric-of-interest for the application, for finite element implementations of each of the three kernel versions (PA, Fused MF, Fused PA) and their respective FP64 tensor-core-accelerated versions (DMMA PA, DMMA Fused MF, DMMA Fused PA). Runtimes were averaged over ten consecutive timesteps (40 operator applications), discarding the initial ten timesteps as warm-up.

The strong scalability results, shown in Figure~\ref{fig:strong-scaling}, demonstrate ideal (linear) strong scaling for all kernels over the first 8$\times$ increase of nodes (from 144 to 1,152 GH200); at full system size (i.e., over a 64$\times$ increase of nodes), the finite element implementations achieve excellent 86--91\% strong parallel efficiencies.
Weak scalability results, depicted in Figure~\ref{fig:weak-scaling}, show that ideal (linear) weak scaling (i.e., $\sim$100\% weak parallel efficiency) is achieved for all of the six kernel implementations over a 64$\times$ increase of nodes.

%% file: 6_related.tex
Since the release of tensor cores in NVIDIA's Volta GPUs~\cite{nvidia-v100}, many HPC applications have successfully leveraged them to improve performance~\cite{8425458}.
With the addition of FP64 tensor cores in NVIDIA's Ampere architectures, adoption expanded further,
enabling applications requiring double-precision computations to benefit as well~\cite{yu2022gpu}.
Early efforts focused on large matrix operations, primarily using vendor-optimized libraries.
Subsequent FP64 tensor core research has diverged into two paths.
For large matrices, the Ozaki scheme~\cite{ootomo2024dgemm,ozaki2025ozaki} emulates FP64 calculations with integer tensor cores, yielding higher performance than FP64 tensor core usage.
For smaller matrices, recent studies~\cite{gu2025sptcstencil,8425458} demonstrate gains from directly employing FP64 tensor cores.

Building on this latter approach, our work extends tensor core acceleration to more complex applications on newer hardware (GH200 and GB200),
while also quantifying energy efficiency improvements and integrating the optimizations into the production MFEM library for a full PDE-based application.
Related finite element studies align with these trends.

For instance, Cui et al.~\cite{cui2024acceleration} explored NVIDIA A100 tensor cores to speed up tensor-product operations in a finite element Poisson solver,
reporting significant gains, but did not include kernel fusion or an energy efficiency analysis.
Relevant work was also presented in the CEED-MS40 report~\cite{ceed_ms40_202} of the Center for Efficient Exascale Discretizations~\cite{ceed},
which examined high-order finite element acceleration using NVIDIA tensor cores and AMD matrix cores for FP64 matrix-matrix multiplications in benchmark kernels on A100 and MI250X GPUs,
reporting speedups over vector instructions mainly for orders above 10.
Our approach builds on these efforts by targeting production-scale, unstructured-mesh operators in a complex tsunami digital twin application,
while leveraging MFEM's modular backends for improved portability across tensor core shapes and precisions.

Complementary efforts have leveraged lower-precision tensor cores to balance performance and accuracy in finite element contexts:
Ruda et al.~\cite{ruda2023fast} used reduced-precision tensor cores on NVIDIA A100 and H100 GPUs for matrix-based methods;
Yamaguchi et al.~\cite{yamaguchi2020low} applied Volta tensor cores to a fast low-order finite element solver for crustal deformation;
and Ichimura et al.~\cite{ichimura2025fast, ichimura2024low} developed explicit structured-mesh wavefield simulations with INT8 tensor cores
to enhance GPU performance and reduce numerical dispersion.

%% file: 7_conclusions.tex
The performance of finite element simulations, including the 2025 Gordon Bell Prize-winning application for real-time tsunami forecasting, relies heavily on small matrix operations.
In this paper, we showed in detail how to further improve the performance of this highly-tuned application using FP64 tensor cores and kernel fusion.
By using tensor cores, we reduced shared memory reads by 4.6$\times$ and increased the performance of key kernels by up to 59\% and energy efficiency by up to 27\% on NVIDIA's GB200 and GH200 chips.
Combined with loop fusion, we obtained speedups of up to 2$\times$ and energy savings of up to 83\%.

The enhancements described in this work were developed within the MFEM library.
The specific FP64 tensor core optimizations and related kernel fusions are currently under review for integration into the public open-source repository~\cite{mfem-web},
which will enable users to leverage these advancements in their own production codes and scientific simulations.